\documentclass{article}
    \usepackage[utf8]{inputenc}
    \usepackage{graphics,amsmath,amssymb,epsfig,epstopdf}
    
    \title{Dark matter fluid constraints from galaxy rotation curves }
    \author{Dalibor Perkovi\'{c}$^{1,}$\thanks{dalibor.perkovic@zvu.hr}  and Hrvoje \v Stefan\v ci\'c$^{2,}$\thanks{hrvoje.stefancic@unicath.hr} }
    
    \date{
    \centering
    $^{1}$ University of Applied Health Sciences, Mlinarska street 38, 10000 Zagreb, Croatia \\
    \vspace{0.2cm}
    $^{2}$ Catholic University of Croatia, Ilica 242, 10000 Zagreb, Croatia }

\begin{document}
    
    \maketitle
    
    \abstract{Galaxy rotation curves are considered to be convincing evidence for dark matter or some dynamically equivalent alternative mechanism. Starting only from the rotation curve data, we present a model independent approach of testing a general hypothesis that dark matter has the properties of a barotropic fluid. It is shown how the speed of sound squared can be expressed in terms of rotation curve data and their radial derivatives and how model independent constraints can be obtained from the requirements that it is confined between 0 and $c^2$. Using the Milky Way rotation curve data available in the literature, we obtain the constraints on the barotropic fluid speed of sound and illustrate the potential of this approach. Technical challenges, limitations and possible future extensions and improvements of the proposed approach are discussed.}
    
    \section{Introduction}
    
    The most intriguing aspect of modern cosmology is the unknown nature of its dark sector. Available observational data (see e.g. \cite{SNIa1,SNIa2,CMB1,CMB2,BAO}), increasing both in quality and quantity, reveal effects in global cosmic expansion, large scale structure formation, galaxy cluster and galaxy dynamics, the explanation of which requires novel cosmic mechanisms. The precise mechanism(s) behind these effects are currently not known and they are sometimes referred to as the cosmic {\em dark sector}. A majority of proposed solutions to late-time accelerated cosmic expansion are either based on an additional cosmic component with a negative pressure, called dark energy (DE) \cite{DERev1,DERev2,DERev3,DERev4,DERev5} or on modifications of gravitational interaction beyond General Relativity \cite{ModGravRev1,ModGravRev2,ModGravRev3}. The effects at the level of cosmic structures are most frequently modeled in terms of additional matter component, called dark matter \cite{DM1,DM2,DM3,DM4,DM5,DM6,DM7}, but alternative mechanisms of modified dynamics, such as MOND and its generalizations \cite{MOND1,MOND2,MOND3,MOND4,MOND5}, have been proposed and are currently being actively investigated. A conceptually simple realization of the cosmic dark sector is used in the presently standard cosmological model, $\Lambda$CDM model, in which the cosmological constant serves as dark energy and the dark matter component is cold dark matter. Despite being simple and in overall accordance with available observational data, tensions in the values of the Hubble constant $H_0$ measured at low and high redshift \cite{H0tension1,H0tension2,H0tension3,H0tension4}, have recently triggered various (re)evaluations of alternative models of the cosmic dark sector.
    
    The research of the dark sector of the universe is in a situation where the fundamental properties of these mechanisms are still not settled e.g. whether they are cosmic components or manifestations of modified gravitational interaction or more generally modified dynamics. Even for the most studied options (such as dark matter or dark energy components) the available data cannot discriminate among numerous proposed models. In such circumstances more generic, model independent tests of the nature of the cosmic dark sector are welcome.
    
    One of most widely used assumptions for cosmic components is that they behave as perfect fluids. An additional, frequently employed requirement is that the perfect fluid is barotropic, i.e. that its pressure is a function of its energy density only, $p=p(\rho)$. In this broad class of models, the fluid speed of sound follows directly form the fluid Equation of State (EoS), $c_s^2=\frac{d \, p}{d \, \rho}$. Numerous recent models of dark matter go beyond the WIMP hypothesis and allow a considerable interaction between DM particles \cite{selfint1,selfint2,selfint3}. Some of these models have the properties of barotropic fluids with a non-negligible pressure. Prominent examples are the Bose Einstein Condensate (BEC) models of dark matter \cite{BEC1,BEC2,BEC3,BEC4,BEC5,BEC6}, with $p \sim \rho^2$ and superfluid dark matter \cite{SF1,SF2,SF3,SF4,SF5} with $p \sim \rho^3$. Given the physical meaning of $c_s^2$, a requirement $0 \le c_s^2 \le c^2$ can be imposed, where $c$ denotes speed of light \footnote{The results obtained in this paper may also be compared with other constraints stemming from specific models of dark matter barotropic fluids. As our main goal is in model independent analyses, we restrict ourselves to this fundamental constraint. }. This research program was systematically applied to parametrizations of dark energy EoS available in the literature in \cite{mi3}. The result of this analysis is that a large proportion of dark energy EoS parametrizations is not compatible with dark energy being a barotropic perfect fluid. It should be stressed that our approach is not applicable to models in which the fluid description is an effective approach and speed of sound, as defined above, may not be physically meaningful.
    
    In this paper we apply the line of reasoning elaborated in \cite{Caplar,mi1,mi2,mi3,mi4} to astrophysical systems. In particular, we develop a model independent framework of testing the hypothesis of dark matter being a barotropic perfect fluid using the galaxy rotation curve observational data.   
    
    The necessary assumptions for the study of astrophysical systems (primarily galaxies) in this paper are that they are spherically symmetric, all components in the system are barotropic fluids, the components are non-interacting and that non-relativistic effects are negligible. Furthermore, it is assumed that rotation curve data of sufficient quality and radial extension are available. These assumptions are clearly idealizations used to simplify the calculations and the comparison with the data. Though unrealistic at smaller radial distances, we will argue that at larger radial distances approximate spherical symmetry may be used for the illustration of our approach.
    
    The main contributions of this paper are the following: We demonstrate how a general concept of dark matter as a barotropic fluid can be tested in a model independent way; The  potential of this approach using Milky Way rotation curve data is illustrated; Some challenges in numerical computation of $c_s^2$ are investigated; The extension of this approach to non-spherical systems and its application to single component systems such as ultra diffuse galaxies is discussed.  
    
    In our considerations we focus our attention to Milky way for several reasons. The rotation curve for Milky Way is available for large radial distances, where the analysis presented in this paper is most adequate for the study of dark matter properties. The measurements of galactic properties important for future refinements of our approach (such as baryonic mass density) are more detailed for Milky Way compared to other galaxies. Finally, as Milky Way is our domicile galaxy, new ways of testing its properties are intrinsically important and interesting.   
    
    The outline of the paper is the following. The first section brings the introduction and an overview of the idea of model independent testing of dark matter being a barotropic fluid. In the second section the formalism of the testing approach is elaborated. In the third section the Milky Way rotation curve data are discussed and the computational methodology is presented. The fourth section brings the results for the Milky Way galaxy. In the final fifth section we discuss the obtained results, outline possible extensions of the presented approach and close the paper with conclusions.

    \section{Fluid dark matter quantities in terms of rotation curve properties}

    The hydrostatic equilibrium of a single relativistic fluid component in a spherically symmetric configuration is given by the famous Tolman-Oppenheimer-Volkoff (TOV) equation
    \begin{equation}
    \label{eq:TOV}
    -r^2 \frac{d \, p(r)}{d \, r}=G {\cal M}(r) \left(\rho(r)+\frac{p(r)}{c^2}\right) \left(1-\frac{2 G {\cal M}}{c^2 r} \right)^{-1} \, ,
    \end{equation}
    where $\rho(r)$ and $p(r)$ correspond to mass density and pressure of the fluid, with ${\cal M}(r)=4 \pi \int_0^r \rho(r') r'^2 d \, r'$ being the mass enclosed within a sphere of radius $r$.
    
    If relativistic effects are negligible, and $\frac{p(r)}{c^2}$ is much smaller than $\rho(r)$, the equilibrium configuration is given by the equation
    \begin{equation}
    \label{eq:equilibrium}
    -r^2 \frac{d \, p(r)}{d \, r}=G {\cal M}(r) \rho(r) \, .
    \end{equation}
    
    If the considered fluid is {\em barotropic} (the fluid pressure is a function of mass density only, $p=p(\rho)$),  using the definition of the speed of sound for a barotropic fluid, $c_s^2=\frac{d \, p}{d \, \rho}$, Eq. (\ref{eq:equilibrium}) can be written as
    \begin{equation}
    \label{eq:equilibrium_2}
    -r^2 c_s^2 \frac{d \, \rho(r)}{d \, r}=G {\cal M}(r) \rho(r) \, .
    \end{equation}
    
    In a non-relativistic Newtonian approximation the circular orbit at radius $r$ of a test particle of mass $m$ is determined by the equation
    \begin{equation}
    \label{eq:circular}
    \frac{m v(r)^2}{r}=G \frac{m {\cal M}(r)}{r^2} \, .
    \end{equation}
    This allows us to express the mass with a sphere of radius $r$ as
    \begin{equation}
    \label{eq:Mofrv}
    {\cal M}(r)=\frac{r v(r)^2}{G} \, .
    \end{equation}
    
    Using the definition of ${\cal M}$, one can obtain the expression for $\rho(r)$ in terms of $v(r)$ and $r$:
    \begin{equation}
    \label{eq:rhoofrv}
    \rho(r)=\frac{1}{4 \pi G r^2} \left(v^2+r \frac{d \, v(r)^2}{d \, r} \right) \, .
    \end{equation}
    Then it is straightforward to obtain the expression for the derivative of energy density
    \begin{equation}
    \label{eq:drhodrofrv}
    \frac{d \, \rho}{d \, r} = \frac{1}{4 \pi G} \left(-\frac{2 v(r)^2}{r^3} + \frac{1}{r} \frac{d^2 \, v(r)^2}{d r^2} \right) \, .
    \end{equation}
    
    After inserting (\ref{eq:Mofrv}), (\ref{eq:rhoofrv}) and (\ref{eq:drhodrofrv}) into (\ref{eq:equilibrium_2}) the expression for the speed of sound as a function of rotation speed and radius then follows:
    \begin{equation}
    \label{eq:cs2ofrv}
    c_s^2=v(r)^2 \frac{v(r)^2+r \frac{d \, v(r)^2}{d \, r}}{2 v(r)^2 - r^2 \frac{d^2 \, v(r)^2}{d r^2}} \, .
    \end{equation}
    
    If we adopt the physical bound $0 \le c_s^2 \le c^2$, then the requirement that matter be a barotropic fluid puts constrains on the shape of the rotation curve $v(r)$. As discussed in more detail in the remainder of the paper, the practical usefulness of this constraint is closely related to data quality.
    
    Next we consider a spherically symmetric configuration with two components which interact only gravitationally. The hydrostatic equilibrium equations for each of components, denoted by subscripts $1$ and $2$ is
    \begin{equation}
    \label{eq:equilibrium_two}
    -r^2 \frac{d \, p_{1,2}(r)}{d \, r}=G {\cal M}(r) \rho_{1,2}(r) \, ,
    \end{equation}
    where ${\cal M}(r)=4 \pi \int_0^r \rho(r') r'^2 d \, r'$ is the mass of both components enclosed in a sphere of radius $r$. Here we also use $\rho(r)=\rho_1(r)+\rho_2(r)$ and ${\cal M}(r)={\cal M}_1(r)+{\cal M}_2(r)$, where ${\cal M}_{1,2}(r)=4 \pi \int_0^r \rho_{1,2}(r') r'^2 d \, r'$.
    
    A circular orbit at radius $r$ of a test particle of mass $m$ is also determined by the equation
    \begin{equation}
    \label{eq:circular_two}
    \frac{m v(r)^2}{r}=G \frac{m {\cal M}(r)}{r^2} \, ,
    \end{equation}
    which leads to expressions
    \begin{equation}
    \label{eq:Mofrv_two}
    {\cal M}(r)={\cal M}_1(r)+{\cal M}_2(r)=\frac{r v(r)^2}{G} \, 
    \end{equation}
    and 
    \begin{equation}
    \label{eq:rhoofrv_two}
    \rho(r)=\rho_1(r)+\rho_2(r)=\frac{1}{4 \pi G r^2} \left(v^2+r \frac{d \, v(r)^2}{d \, r} \right) \, .
    \end{equation}
    
    Combining both equations given in (\ref{eq:equilibrium_two}), one easily obtains 
    \begin{equation}
    \label{eq:equilibrium_two2}
    -r^2\left( c_{s,1}^2 \frac{d \, \rho_1(r)}{d \, r}+  c_{s,2}^2 \frac{d \, \rho_2(r)}{d \, r}\right)=G {\cal M}(r) \rho(r) \, .
    \end{equation}
    The right-hand side of this equation can be readily expressed in terms of $v(r)^2$ and $r$ using (\ref{eq:Mofrv_two}) and (\ref{eq:rhoofrv_two}). 
    
    A considerable simplification of (\ref{eq:equilibrium_two2}) is obtained if there is a region of large $r$ where $\rho_2(r)$ and $p_2(r)$ are negligible. Then in this region of sufficiently large $r$ the expression (\ref{eq:equilibrium_two2}) takes the form
    \begin{equation}
    \label{eq:equilibrium_two2_simple}
    -r^2 c_{s,1}^2 \frac{d \, \rho_1(r)}{d \, r}=G ({\cal M}_1(r)+{\cal M}_2) \rho_1(r) \, .
    \end{equation}
    Here ${\cal M}_2$ is the total mass of the component $2$ in the system. In this regime the expressions for $\rho_1$ and $c_{s,1}^2$ are given as functions of orbital velocity and ts radial derivatives by expressions (\ref{eq:rhoofrv}) and (\ref{eq:cs2ofrv}), respectively.
    
    For a two component system, even more realistic testing of dark matter properties can be made if additional information on one of components can be independently obtained. A particular instance of this situation is if baryonic matter distribution is (precisely) known from direct observations (other that rotation curves). Namely, for a rotation curve velocity one can write
    \begin{equation}
    \label{eq:twov}    
        v^2=v_1^2+v_2^2 \, ,
    \end{equation}
    where $v_{1,2}^2=\frac{G {\cal M}_{1,2}}{r}$. If $v_2^2$ can be calculated from additionally available information (e.g. from known matter distribution), $v_1^2$ can be calculated from measured $v^2$ and calculated $v_2^2$ as $v_1^2=v^2-v_2^2$. Then it is straightforward to obtain
    \begin{equation}
    \label{eq:cs2_v_v1}
    c_{s,1}^2=v(r)^2 \frac{v_1(r)^2+r \frac{d \, v_1(r)^2}{d \, r}}{2 v_1(r)^2 - r^2 \frac{d^2 \, v_1(r)^2}{d r^2}} \, .
    \end{equation}

    The developed formalism presented above allows for testing several hypotheses using rotation curve data only. The first one starts from the requirement that energy density should not be negative, i.e. $\rho \ge 0$ which can be tested using (\ref{eq:rhoofrv}). Such attempts of reconstructing the energy density from the rotation curve data have already been described in the literature (see e.g. \cite{BT}). Finding statistically significant negative values of $\rho$ may point to some alternative mechanism behind rotation curve dynamics other than dark matter as a barotropic fluid. A hypothesis that matter producing rotation curve dynamics behaves as a barotropic fluid can be tested using expression (\ref{eq:cs2ofrv}), where we are interested if the null hypothesis of $0 \le c_s^2 \le c^2$ can be rejected. In particular, it can be studied if statistically significant negative values of $c_s^2$ can be found at some radial distance values. Should empirical rotation curves reveal statistically significant negative values of $c_s^2$, the component producing the rotation curve dynamics cannot be a stable barotropic fluid, i.e. it should exhibit instabilities. An additional advantage of this approach is that the speed of sound expression is valid for any $r$ so this provides an opportunity to test the requirement for $c_s^2$ for a range of $r$ i.e. for an entire rotation curve.   
    
    Apart from testing the aforementioned hypotheses, additional interesting information can be achieved using the rotation curve data. Direct integration of (\ref{eq:equilibrium}) yields an expression for the pressure as a function of radial distance:
    \begin{equation}
    \label{eq:pofrv}
    p(r)=p_0 - \int_{r_0}^r \frac{v(r')^2}{4 \pi G r'^3} \left(v(r')^2+r' \frac{d \, v(r')^2}{d \, r'} \right) d \, r'\, ,
    \end{equation}
    where the subscript $0$ refers to some arbitrary rotation curve point. This result shows that if the value of pressure at some radial distance $r_0$ is known, the pressure function $p(r)$ can be reconstructed from the rotation curve data. The expression (\ref{eq:pofrv}) can be readily integrated by parts which results in an expression for $p(r)$ without derivatives of rotation curve speed function:
    \begin{equation}
    \label{eq:pwithoutderivative}
    p(r)=p_0 - \frac{1}{8 \pi G} \left[ \frac{v(r)^4}{r^2} - \frac{v(r_0)^4}{r_0^2} + 4 \int_{r_0}^r \frac{v(r')^4}{r'^3} d \, r' \right] \, .
    \end{equation}
    Compared to expressions for mass density and speed of sound, the expression for pressure given in (\ref{eq:pwithoutderivative}) is more stable since it does not require calculation of orbital speed derivatives. However, it is dependent on the value $p_0$ which has to be determined by some other method. Once the pressure function is known, it is straightforward to obtain the parameter of Equation of State $w(r)=p(r)/\rho(r)$. Finally, one can plot the $(c_s^2,w)$ relation to obtain a deeper insight into the nature of the matter component producing rotation curve dynamics in the $c_s^2(w)$ formalism \cite{mi1,mi2,mi3,mi4}. Spherical symmetry assumed in this section may not be adequate for the description of some systems, such as galaxies at smaller radial distances. However, in the following section we show in which circumstances certain systems can have approximate spherical symmetry.   
    
    
    \section{Data and computational methods}
    
    \subsection{Milky Way data}
    
    As an illustration of the proposed approach we apply the developed testing scheme to the data publicly available in Ref. \cite{MW}. In this reference a rotation curve for the Milky Way is constructed for radial distances from the galactic center in the range from $\sim 0.2$ kpc to $\sim 200$ kpc. As stated in \cite{MW}, datasets for several different types of tracer objects, both disk tracers and non-disk tracers, need to be combined in a common methodological framework to produce a rotation curve with such a large radial range. Furthermore, the construction of the rotation curve of the Milky Way in \cite{MW} is made without assuming any specific dark matter model or halo structure model. These features of rotation curve construction are particularly convenient for the analysis presented in this paper since: i) a single rotation curve dataset is available and there is no need to work with multiple, possibly quite disparate, observational datasets, ii) the dataset extends to very large radial values ($\sim200$ kpc) where the conditions for testing dark matter properties are the best and influence of the baryonic matter is expected to be small and iii) we aim at model independent testing of general properties of dark matter what also requires data obtained in a model independent way.   
    In our testing if the dark matter can have properties of a barotropic fluid, in this paper we confine our analysis to radial distances above 28 kpc. In this way we avoid the region of radial distances where peculiarities of galactic baryon distribution might significantly interfere with the applicability of simple expressions such as (\ref{eq:rhoofrv}) and (\ref{eq:cs2ofrv}). Namely, we assume that the dark matter halo is spherical and with increasing radial distance the influence of non-spherical baryonic distribution is progressively smaller. In this way at sufficiently large radii the system may be taken as approximately spherically symmetric. In this paper we assume that 28 kpc is the lower limit of this region of approximate spherical symmetry. Owing to large interval of radial distances for which measurements of orbital distances are available in \cite{MW}, even such a restricted dataset is still sufficiently large for systematic application of the proposed test if dark matter has the properties of a barotropic fluid.

    \subsection{Computational methods}
    
    The aim of this paper is to compare the values of physical quantities such as $\rho$ and $c_s^2$ calculated from the galaxy rotation curve measurements with the fundamental hypotheses such as $\rho \ge 0$ and $0 \le c_s^2 \le c^2$. Statistical testing of these hypotheses requires knowing the scope of systematic uncertainties, mainly related to numerical calculation of derivatives, and statistical distributions of calculated quantities such as $\rho$ and $c_s^2$ coming from statistical uncertainties of galaxy rotation curves, $v(r)$.     
    
    The main challenge in using expressions like (\ref{eq:rhoofrv}) and (\ref{eq:cs2ofrv}) to calculate observable quantities such as $\rho$ and $c_s^2$ lies in the numerical calculation of first and second spatial derivative of the $v(r)$ function. It is important to assess how sensitive the results for $\rho$, $c_s^2$ and other relevant quantities are to the method of calculation of the said derivatives. Furthermore, as the rotation curve data measurements have some associated uncertainties, an essential question that needs to be addressed is how the uncertainties in the $v(r)$ functions are translated into distributions of calculated values such as $\rho$ and $c_s^2$. We approach this issue from several directions and propose methodologies of quantifying dispersion of the calculated physical values.  We employ several schemes of calculating derivatives to estimate the level of robustness of our results.
    
    For the purpose of numerical processing of (\ref{eq:rhoofrv}) and (\ref{eq:cs2ofrv}), we use two approaches relying on second-order and fourth order Lagrange interpolation polynomials. In the first approach $v^2(r)$ around $r_i$ is represented by a second-order Lagrange interpolation polynomial with three points:
    
    \begin{eqnarray}
    \label{eq:v2Lagrange}
    v^2(r) &=& \frac{(r-r_i)(r-r_{i+1})}{(r_ {i-1}-r_i)(r_{i-1}-r_{i+1})}v_{i-1}^2 
    + \frac{(r-r_{i-1})(r-r_{i+1})}{(r_ {i}-r_{i-1})(r_{i}-r_{i+1})}v_{i}^2 + \nonumber \\
    &+& \frac{(r-r_{i-1})(r-r_{i})}{(r_ {i+1}-r_{i-1})(r_{i+1}-r_{i})}v_{i+1}^2 \, .
    \end{eqnarray}
    
    For the fourth-order Lagrange interpolation polynomial with five points the expression for $v^2(r)$ around $r_i$ is:
    
    \begin{eqnarray}
    \label{eq:v4Lagrange}
    v^2(r) &=& \frac{(r-r_{i-2})(r-r_{i-1})(r-r_{i})(r-r_{i+1})}{(r_{i+2}-r_{i-2})(r_{i+2}-r_{i-1})(r_{i+2}-r_{i})(r_{i+2}-r_{i+1})}v^2_{i+2} \nonumber \\
    &+&\frac{(r-r_{i-2})(r-r_{i-1})(r-r_{i})(r-r_{i+2})}{(r_{i+1}-r_{i-2})(r_{i+1}-r_{i-1})(r_{i+1}-r_{i})(r_{i+1}-r_{i+2})}v^2_{i+1} \nonumber \\
    &+& \frac{(r-r_{i-2})(r-r_{i-1})(r-r_{i+1})(r-r_{i+2})}{(r_{i}-r_{i-2})(r_{i}-r_{i-1})(r_{i}-r_{i+1})(r_{i}-r_{i+2})}v^2_{i}\nonumber \\
    &+&\frac{(r-r_{i-2})(r-r_{i})(r-r_{i+1})(r-r_{i+2})}{(r_{i-1}-r_{i-2})(r_{i-1}-r_{i})(r_{i-1}-r_{i+1})(r_{i-1}-r_{i+2})}v^2_{i-1} \nonumber \\
    &+&\frac{(r-r_{i-1})(r-r_{i})(r-r_{i+1})(r-r_{i+2})}{(r_{i-2}-r_{i-1})(r_{i-2}-r_{i})(r_{i-2}-r_{i+1})(r_{i-2}-r_{i+2})}v^2_{i-2} \, .
    \end{eqnarray}
    
    All required derivatives are obtained by analytical differentiation of expressions (\ref{eq:v2Lagrange}) and (\ref{eq:v4Lagrange}).
    
    To assess the importance of local details in the rotation curve (as opposed to the average trend of the rotation curve), rotation curve data were also smoothed using 
    %
    %
    weighted moving average with a moving window whose width was 20 kpc. To create a new smooth velocity curve, a point was created for each kpc, using all available measured data within 10 kpc on each side. The statistical weight of the measured data was linear with the distance of the measured data from the newly constructed point. For example, a point 10 kpc from the newly constructed point or further away had zero statistical weight, a point at exactly the same radius as a newly constructed point had a maximum statistical weight, while a point 5 kpc away from the newly constructed point had half of the maximum  statistical weight.
    
    
    Finally, for each of four chosen approaches (three point and five point, without and with smoothing), an ensamble of values for $\rho$ and $c_s^2$ is calculated using the following procedure. For a three-point approach the required quantities are calculated for all radial values except the first and the last one. For a five-point approach the required quantities are calculated for all radial values except the first two and the last two. The required values for the speed triplet in (\ref{eq:v2Lagrange}) (quintuplet in (\ref{eq:v4Lagrange})) are drawn from normal distributions with expected value and standard deviation from dataset available in \cite{MW}. The values in the ensamble are sorted and boundaries of 68\%, 95\% and 99\% confidence intervals are determined.

    
    \section{Milky Way results}
    
    In this section we present results obtained from the Milky Way rotation curve data described in the preceding section. We start with a general observation that for the flat part of the rotation curve, characterized by $v=const$, from (\ref{eq:cs2ofrv}) it follows $c_s^2=v^2/2 \equiv \alpha = const$. In the flat part of the rotation curve the EoS of the dark matter fluid is then $p=\alpha \rho + \beta$. As an illustration of the order of magnitude for the parameter $\alpha$, its value for the Milky Way rotation curve data, assuming the flat part for the studied area with $r_0 > 28 \; kpc$, is $\alpha=(2.3  \pm 0.1 )\cdot 10^{10} m^2/s^2$. As values of the rotation speed for the flat part of rotation curve in a large majority of spiral galaxies is in the range from 100 km/s to 300 km/s, the corresponding values for $\alpha$ for spiral galaxies are of the order of $10^{10} m^2/s^2$. 
    
    We further organize our results in four subsections. In the first subsection we present results for radial derivatives of rotation curve orbital speeds and compare various techniques of their calculation. In the second subsection we present results for the matter density while the third subsection contains results for the speed of sound squared. Finally, in the fourth subsection we bring results for the dark matter fluid pressure. 
    
    \subsection{Derivatives}
    
    Extracting the first and the second derivatives of $v^2$ with respect to radial distance represents the most challenging part of calculating physical quantities such as mass density $\rho$ and speed of sound squared $c_s^2$. We use two schemes for the calculation of derivatives, based on three point and five point Lagrange interpolation polynomials and apply each of these schemes to original and smoothed data.
    
    \begin{figure}
    \centering
    \includegraphics[scale=0.45]{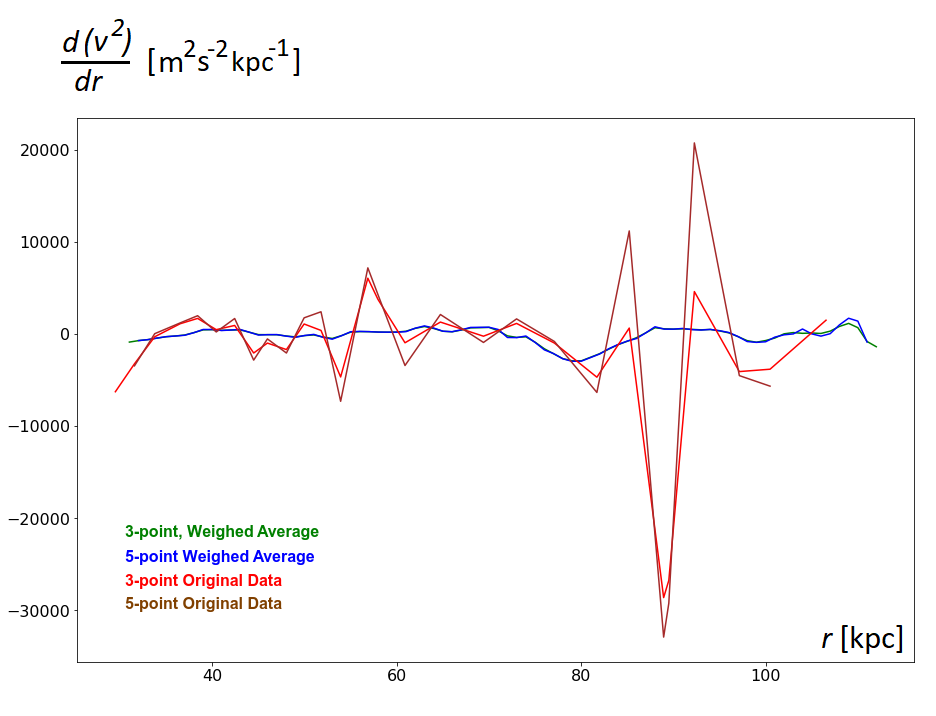}
    \caption{First radial derivatives of orbital speed squared $\frac{d \, v^2}{d r}$, calculated using the three-point scheme for the calculation of derivatives (\ref{eq:v2Lagrange}) and the five-point scheme for the calculation of derivatives (\ref{eq:v4Lagrange}), with smoothing (referred to as Weighted average) and without smoothing (referred to as Original data).  
    }
    \label{fig_d3-d5-Orig-WeighAvg}
    \end{figure}
    
    \begin{figure}
    \centering
    \includegraphics[scale=0.45]{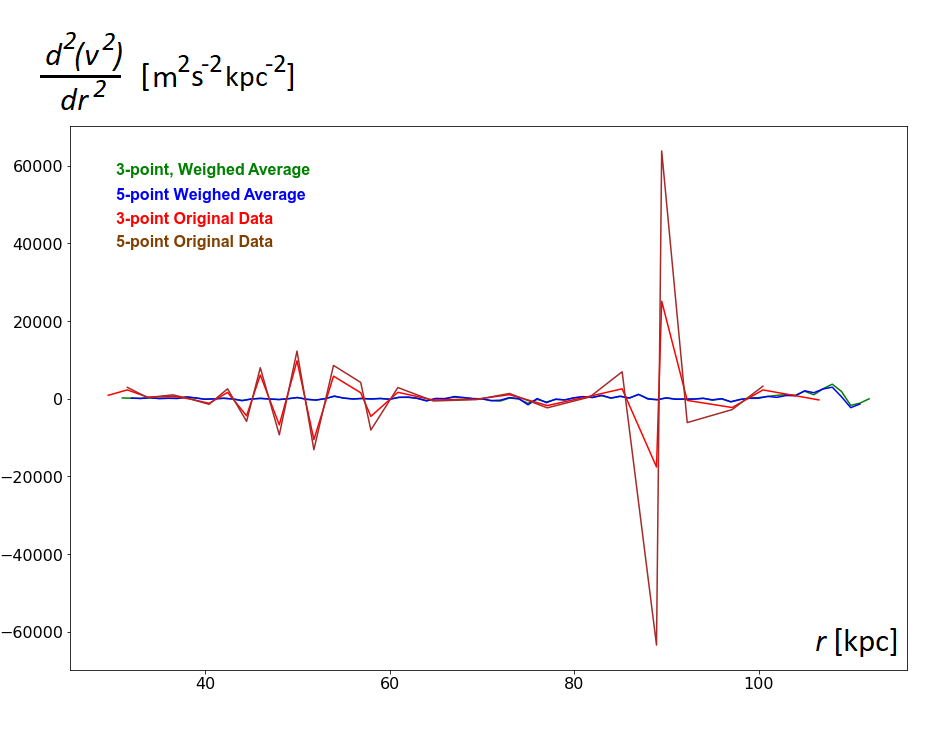}
    \caption{Second radial derivatives of orbital speed squared $\frac{d^2 \, v^2}{d r^2}$, calculated using the three-point scheme for the calculation of derivatives (\ref{eq:v2Lagrange}) and the five-point scheme for the calculation of derivatives (\ref{eq:v4Lagrange}), with smoothing (referred to as Weighted average) and without smoothing (referred to as Original data).
    }
    \label{fig_dd3-dd5-Orig-WeighAvg}
    \end{figure}
    
    The results for the first derivative are presented in Fig. \ref{fig_d3-d5-Orig-WeighAvg}, whereas the results for the second derivative are presented in Fig. \ref{fig_dd3-dd5-Orig-WeighAvg}.
    The values of the first and the second radial derivative of $v^2$, presented in Figures  \ref{fig_d3-d5-Orig-WeighAvg} and \ref{fig_dd3-dd5-Orig-WeighAvg}, exhibit two important features. In the first place, the results obtained for original data (without smoothing) differ significantly from the results obtained for the smoothed data. This result is somewhat expected, given that the derivatives are calculated numerically. Secondly, the differences between results obtained using three-point and five-point schemes (both for original and smoothed data), though they exist, for most radial distances are not large. In particular, there is no difference in sign of derivatives at any radial value for both first and second derivatives. 
    
    Since important information for testing of hypotheses may be obtained from just a subset of radial values where the quantities of interest are calculated, smoothing of the original data may be unjustified. In the remainder of the paper we focus on the results for the original data.
    
    \subsection{Matter density}
    
\begin{figure}
    \centering
    \includegraphics[scale=0.4]{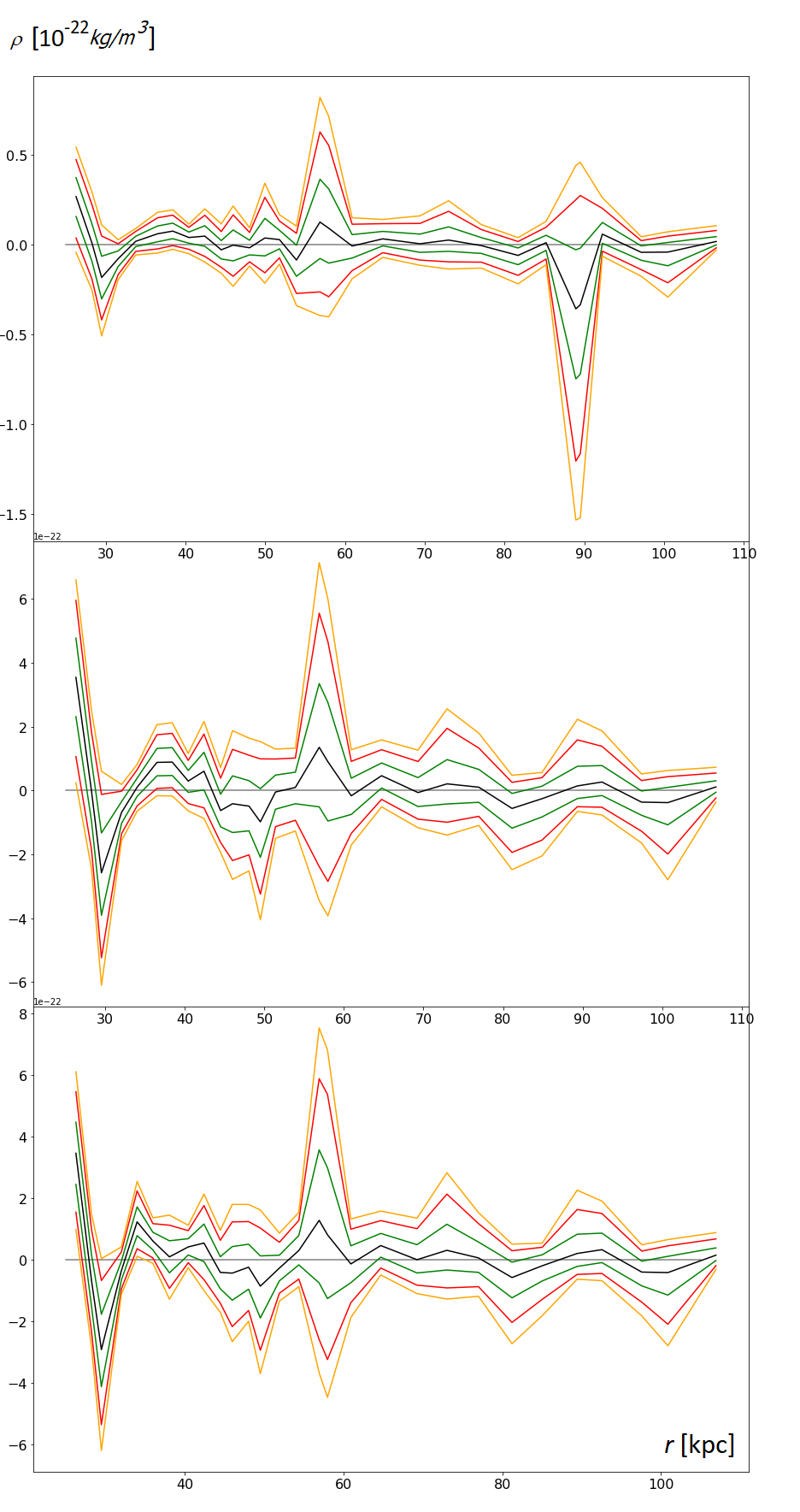}
    \caption{Median (black line) and lower and upper bounds of 68\% (green lines), 95\% (red lines) and 99\% (orange lines) confidence intervals of $\rho$ calculated using three-point scheme for the calculation of derivatives for three combinations of distance of Sun from the galactic center and Sun orbital velocity around the galactic center: $(R_0=8.0$ kpc, $v_0=200$ km/s) (top), $(R_0=8.3$ kpc, $v_0=244$ km/s) (middle) and $(R_0=8.5$ kpc, $v_0=220$ km/s) (bottom). Even though it was declared that we have confined the analysis to radial distances above 28 kpc, a few points below this limit are drawn in these graphs in order to show the downward trend of $\rho$ at lower radii. 
    }
    \label{fig_rho_3point}
    \end{figure}

\begin{figure}
    \centering
    \includegraphics[scale=0.4]{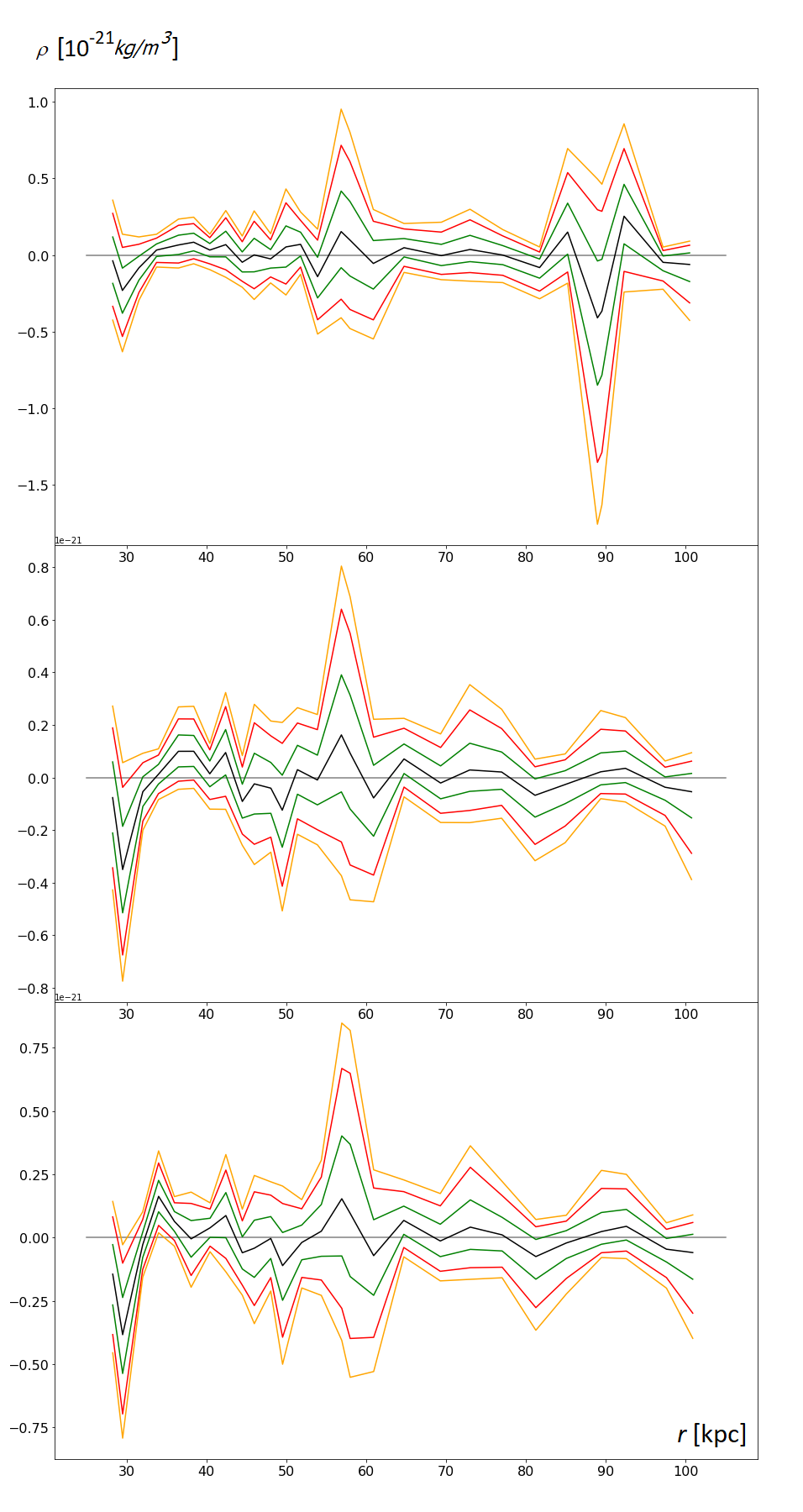}
    \caption{Median (black line) and lower and upper bounds of 68\% (green lines), 95\% (red lines) and 99\% (orange lines) confidence intervals of $\rho$ calculated using five-point scheme for the calculation of derivatives for three combinations of distance of Sun from the galactic center and Sun orbital velocity around the galactic center: $(R_0=8.0$ kpc, $v_0=200$ km/s) (top), $(R_0=8.3$ kpc, $v_0=244$ km/s) (middle) and $(R_0=8.5$ kpc, $v_0=220$ km/s) (bottom). Even though it was declared that we have confined the analysis to radial distances above 28 kpc, a few points below this limit are drawn in these graphs in order to show the downward trend of $\rho$ at lower radii.
    }
    \label{fig_rho_5point}
    \end{figure}

    The matter density distribution is calculated using expression (\ref{eq:rhoofrv}) which employs the first radial derivative of the rotation curve orbital speed. The results for the matter density are presented in Figures \ref{fig_rho_3point} and \ref{fig_rho_5point}. In both figures no smoothing is applied. Both figures consist of three separate plots for different assumptions on Earth radial distance from the galactic center and its orbital speed, as presented in \cite{MW}. Both figures present median (black line) and lower and upper bound of 68\% (green lines), 95\% (red lines) and 99\% (orange lines) confidence intervals. In Fig. \ref{fig_rho_3point}  results for the three-point scheme for the calculation of derivatives (\ref{eq:v2Lagrange}) are presented, whereas in Fig. \ref{fig_rho_5point} results for the five-point scheme for the calculation of derivatives (\ref{eq:v4Lagrange}) are displayed.
    
    
    \subsection{Speed of sound}
    
    \begin{figure}
    \centering
    \includegraphics[scale=0.45]{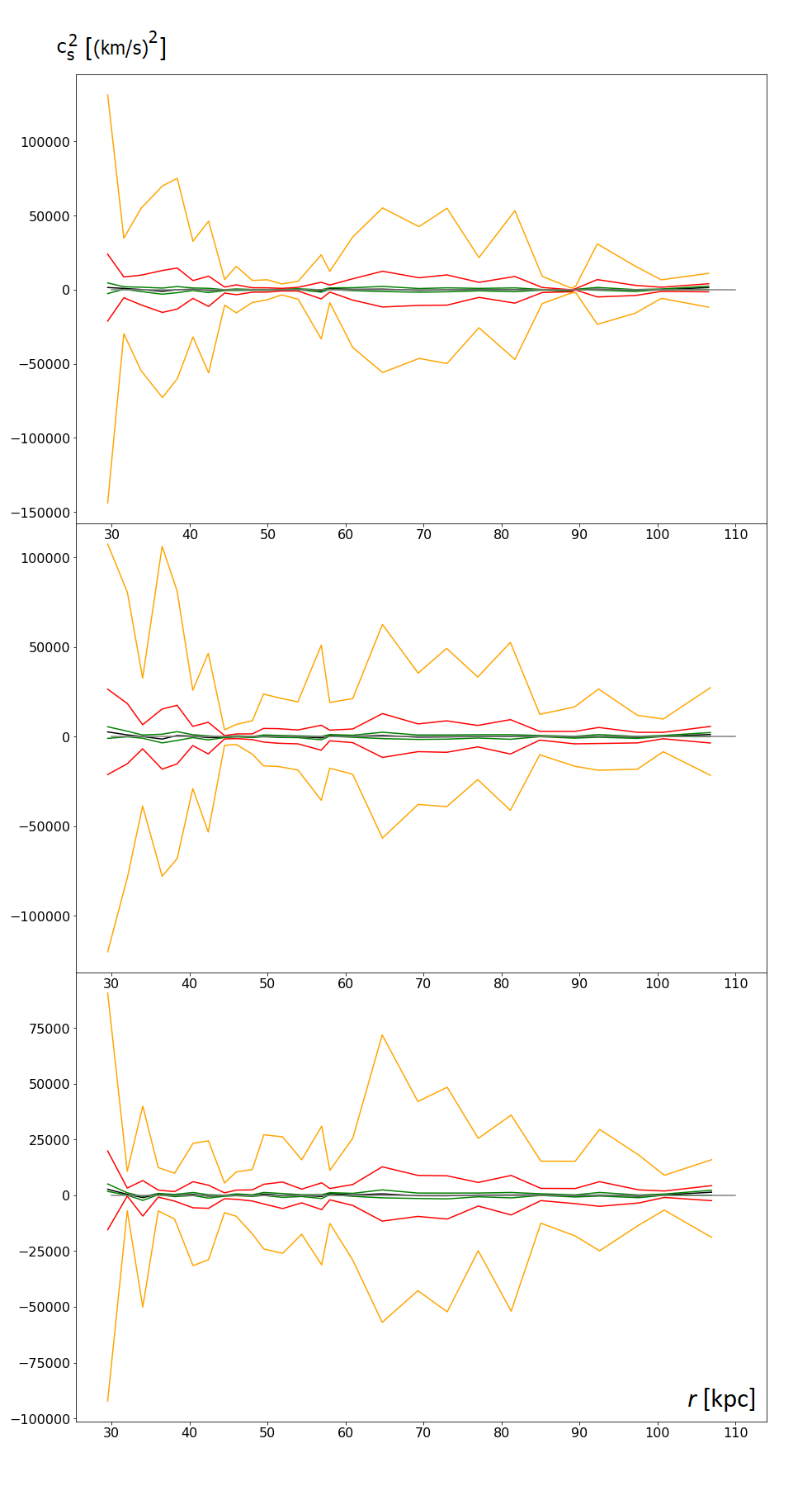}
    \caption{Median (black line) and lower and upper bounds of 68\% (green lines), 95\%  (red lines) and 99\% (orange lines) confidence intervals of $c_s^2$ calculated using three-point scheme for the calculation of derivatives for three combinations of distance of Sun from the galactic center and Sun orbital velocity around the galactic center: $(R_0=8.0$ kpc, $v_0=200$ km/s) (top), $(R_0=8.3$ kpc, $v_0=244$ km/s) (middle) and $(R_0=8.5$ kpc, $v_0=220$ km/s) (bottom).
    }
    \label{fig_cs2_3point}
    \end{figure}
    
    \begin{figure}
    \centering
    \includegraphics[scale=0.45]{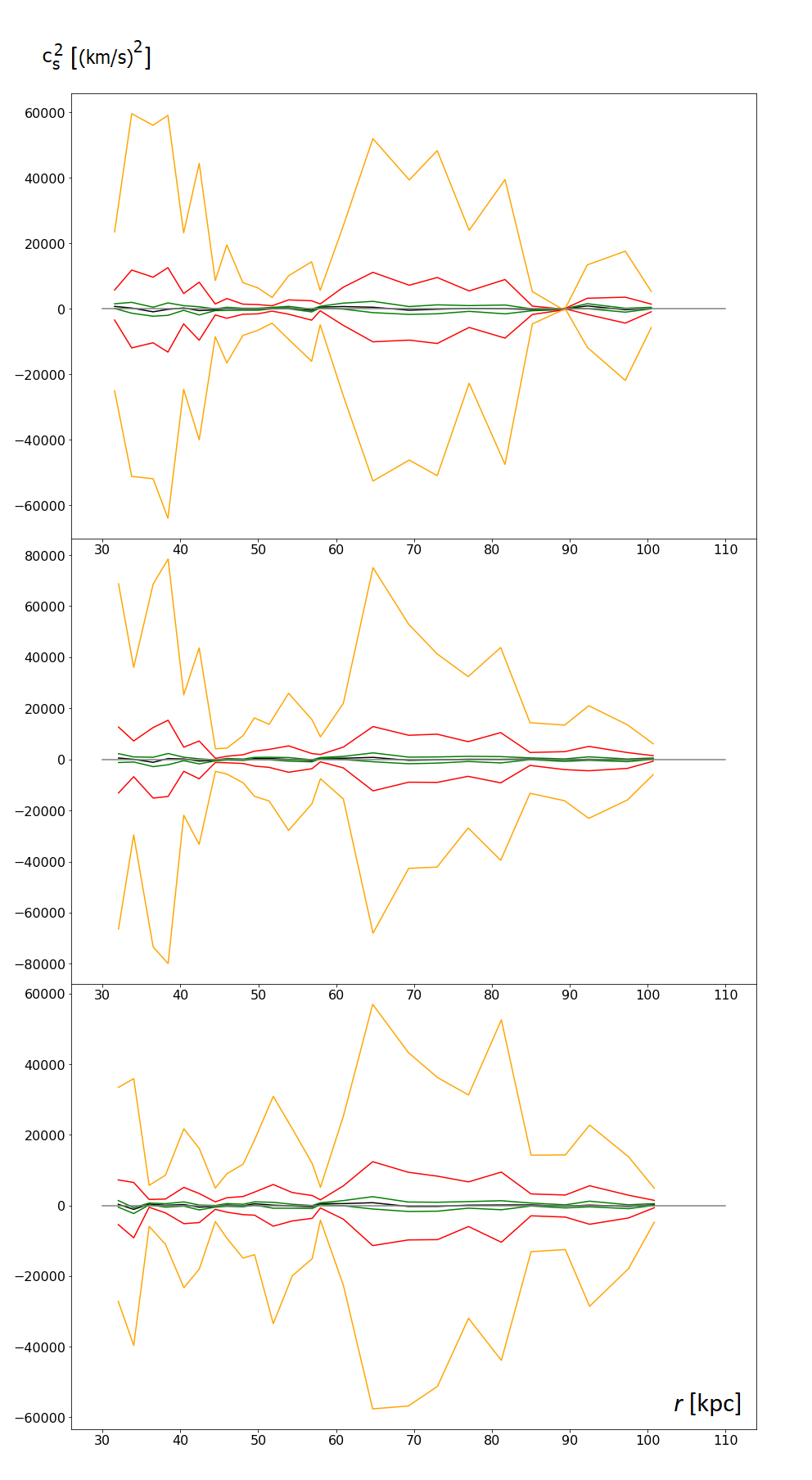}
    \caption{Median (black line) and lower and upper bounds of 68\% (green lines), 95\% (red lines) and 99\% (orange lines) confidence intervals of $c_s^2$ calculated using five-point scheme for the calculation of derivatives for three combinations of distance of Sun from the galactic center and Sun orbital velocity around the galactic center: $(R_0=8.0$ kpc, $v_0=200$ km/s) (top), $(R_0=8.3$ kpc, $v_0=244$ km/s) (middle) and $(R_0=8.5$ kpc, $v_0=220$ km/s) (bottom).
    }
    \label{fig_cs2_5point}
    \end{figure}
    
    \begin{figure}
    \centering
    \includegraphics[scale=0.45]{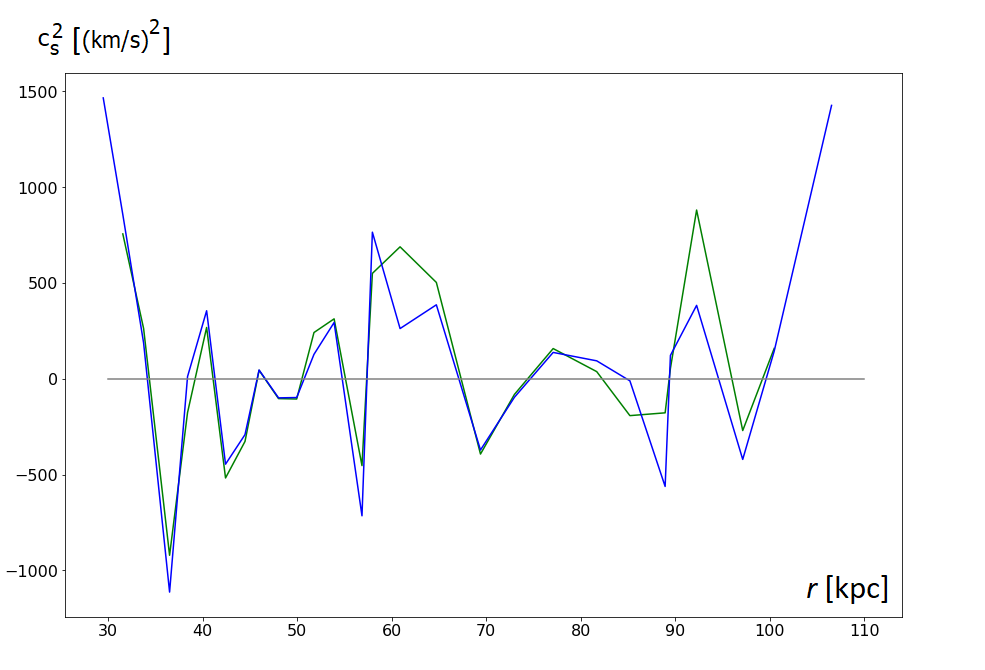}
    \caption{Medians of $c_s^2$ calculated using three-point (blue line) and five-point scheme (green line) for the calculation of derivatives, for the distance of Sun from the galactic center $R_0=8.0$ kpc and its orbital velocity around the galactic center $v_0=200$ km/s.
    }
    \label{fig_cs2median_3point_vs_5point}
    \end{figure}
    
    The results for the speed of sound squared are given in Figures \ref{fig_cs2_3point} and \ref{fig_cs2_5point}. Both figures contain separate plots for three assumptions on Earth radial distance from the galactic center and its orbital speed, as explained in \cite{MW}. In both figures median (black line) and lower and upper bound of 68\% (green lines), 95\% (red lines) and 99\% (orange lines) confidence intervals are displayed and no smoothing is applied. Figure \ref{fig_cs2_3point} contains results for the three-point scheme for the calculation of derivatives (\ref{eq:v2Lagrange}), whereas Figure \ref{fig_cs2_5point} brings results for the five-point scheme for the calculation of derivatives (\ref{eq:v4Lagrange}). Finally, in Fig. \ref{fig_cs2median_3point_vs_5point} the medians for $c_s^2$ calculated using (\ref{eq:v2Lagrange}) and  (\ref{eq:v4Lagrange}) are presented as an illustration of the robustness of calculations to the choice of numerical scheme for the calculation of derivatives.

    \subsection{Pressure}
    
    \begin{figure}
    \centering
    \includegraphics[scale=0.27]{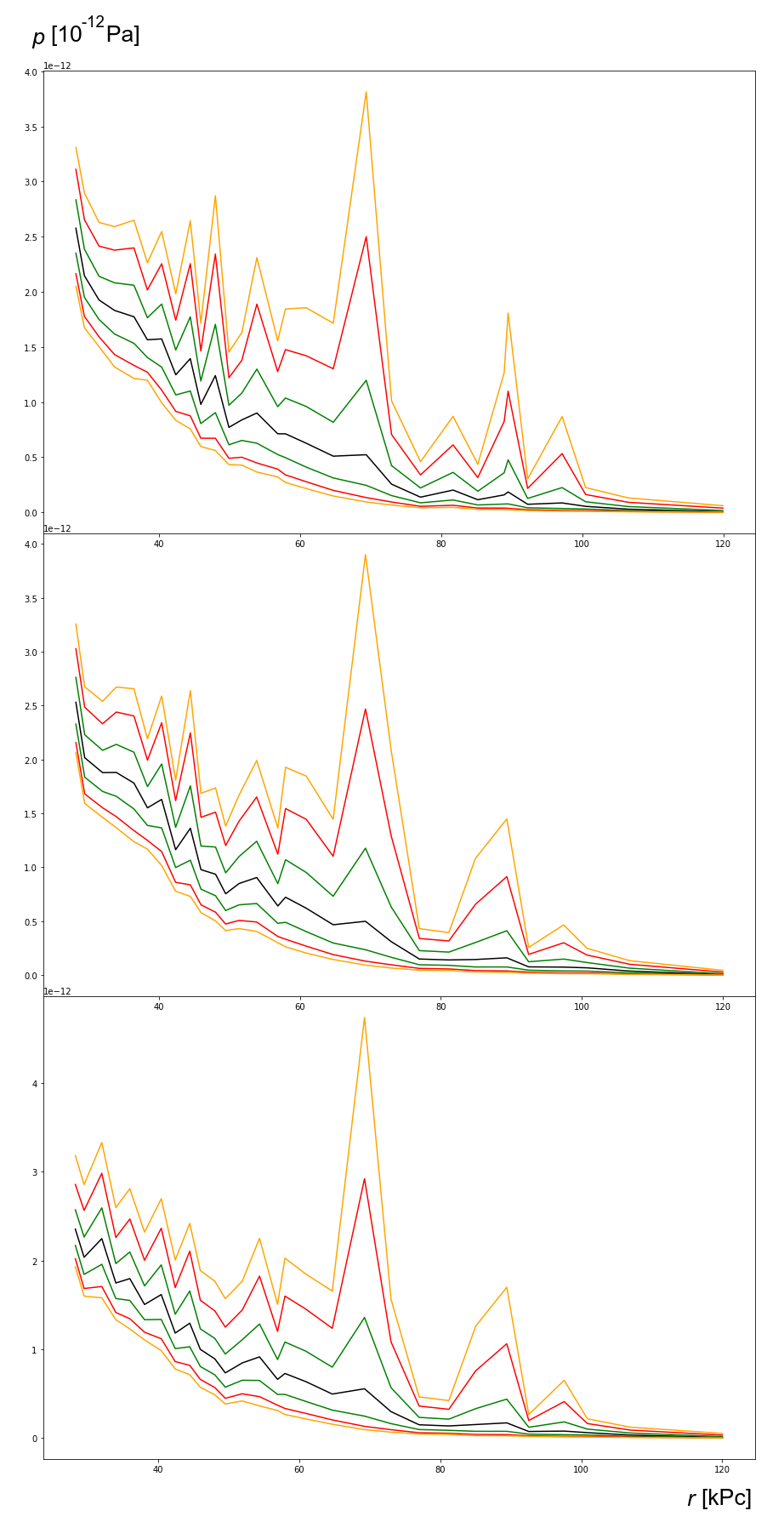}
    \caption{Median (black line) and lower and upper bounds of 68\% (green lines), 95\% (red lines) and 99\% (orange lines) confidence intervals of $p-p_{0}-\frac{v(r_0)^4}{8 \pi G r_0^2}$ for three combinations of distance of Sun from the galactic center and Sun orbital velocity around the galactic center: $(R_0=8.0$ kpc, $v_0=200$ km/s) (top), $(R_0=8.3$ kpc, $v_0=244$ km/s) (middle) and $(R_0=8.5$ kpc, $v_0=220$ km/s) (bottom).
    }
    \label{fig_tlak_x3}
    \end{figure}

    The pressure of dark matter fluid is calculated using the expression (\ref{eq:pwithoutderivative}). As already mentioned, the pressure is defined up to a constant. The results for $p-p_{0}-\frac{v(r_0)^4}{8 \pi G r_0^2}$, for three assumptions on Earth radial distance from the galactic center and its orbital speed, are presented in Fig. \ref{fig_tlak_x3}. The results depicted in Fig. \ref{fig_tlak_x3} were obtained by randomly selecting 1000 sets of values of orbital speed from corresponding normal distributions ${\cal N}(v_1, \sigma_i)$ at each radial value $r_i$. The radial value $r_0$ is chosen as the largest radial value in the dataset. Median of $p-p_{0}-\frac{v(r_0)^4}{8 \pi G r_0^2}$ is denoted by a black line, whereas intervals of 68\%, 95\% and 99\% confidence intervals are given in green, red and orange lines, respectively.

    \section{Discussion and conclusions}
    
    The results presented in the preceding section should  primarily serve as a pilot study of the proposed approach to the testing of dark matter as a barotropic fluid. In the interpretation of the obtained results one should keep in mind that they were obtained using the assumption of spherical symmetry which is only an approximation increasingly more accurate with the increase of radial distance.
      
    For a concrete barotropic fluid model of dark matter with an EoS $p=p(\rho)$, it is possible to perform fits to rotation curve data and obtain estimates of EoS parameters. Some examples of such analyses are \cite{SF5} for the superfluid dark matter in Milky Way or \cite{BEC6} for the BEC dark matter in the SPARC sample. In this paper, we focus on model independent analysis and report results which can be obtained without any additional assumption on the form of dark matter fluid EoS. However, it would be very interesting to combine the model independent results obtained here with the analyses of specific DM models. This interesting possibility is left for future considerations.
    
    One of the most important results for $\rho$ and $c_s^2$ quantities is that at some radii their values are below 0 at $68\%$ or even $95\%$ and, in very rare cases, $99\%$ confidence levels. Therefore, it is reasonable to expect that with better (more accurate) data, larger simulation datasets and better evaluation of the baryonic contribution this approach might provide stringent test of dark matter as a barotropic fluid.  Furthermore, statistically significant negative values of $c_s^2$ at some radii might provide information on regions of possible dark matter instability. Beside interest in this fascinating possibility in itself, it could also provide valuable insight into complex dynamics of galactic structures.
    
    As already stated, the obtained results can be more reliably used in testing the nature of dark matter at larger radii where the baryonic component density is negligible and the baryonic contribution can be accurately approximated as if all baryonic mass were concentrated in the galactic center. Furthermore, with the growth of $r$ the assumption of spherical symmetry is an increasingly better approximation. For this reason we restrict our analysis to Milky Way rotation data with radial distance from the galactic center larger than 28 kpc. 
    
    The single-component approach used in this paper could, in principle, be of direct interest in modelling systems consisting predominantly of a single component such as in ultradiffuse galaxies \cite{UDG1,UDG2,UDG3}.   
    
    As already demonstrated in (\ref{eq:cs2_v_v1}), more precise testing for $c_s^2$ can be obtained if the contribution of the baryonic component to $v^2(r)$ is subtracted from the measured $v^2(r)$ data. Such a step requires detailed knowledge of galactic baryonic matter distribution and numerical calculation of the corresponding gravitational potential (or force). In that case, it is neccessary to use axisymmetric models for baryonic component or in cases of spiral galaxies even non-axisymmetric models or representations, see e.g. \cite{BT}.  After this subtraction, using (\ref{eq:cs2_v_v1}) testing of fundamental properties of $c_s^2$ can be made at all available values of $r$.   
    
    The usefulness of the proposed approach is clearly limited by the quality of the galaxy rotation curve data. The rotation speed need to be measured at more densely spaced values of the radial distance $r$, with smaller errors i.e. uncertainties, and at as broad interval of $r$ as possible. A more detailed analysis using present and future observational programs in Milky Way and other galaxies and more refined methods of calculation can be reasonably expected to provide more stringent observational constraints on the nature of dark matter and possibly (dis)prove the hypothesis of dark matter behaving like a barotropic fluid.       
    
    In conclusion, rotation curves of spiral galaxies provide a model independent approach to testing of the  dark matter concept. If the dark matter component can be represented as a barotropic perfect fluid, the expressions for physical quantities of the dark matter component, such as matter density and speed of sound, can be obtained using the rotation curve data only. Whereas reconstruction of the dark matter density has been considered in the literature before, determination of speed of sound of the dark matter barotropic fluid from the rotation curve data is a novelty introduced, to the best of our knowledge, for the first time in this paper. The biggest systematic challenge is the way radial derivatives of the rotation speed are numerically calculated and different schemes of calculation have been used to assess the sensitivity of results on the calculation scheme used. The main goal of this paper is, apart from the introduction of this novel approach, to estimate if the approach can be usefully applied using the observational Milky Way rotation curve data available in the literature. With this goal in mind, we perform a simplified analysis which neglects some details of galactic structure and its results describe dark matter properties only at larger distances from the center of the galaxy. The results obtained for the Milky Way galaxy show that using the chosen dataset and the simplified analysis there is hardly any radial distance at which $\rho$ or $c_s^2$ would be negative at a $99\%$ confidence level. With the more precise modeling of all contributions to rotation curve dynamics and more precise data, the approach presented in this paper could become a promising venue for model independent testing of dark matter properties. Such refined analyses represent an important challenge for future work.

    \end{document}